\documentclass[copyright]{eptcs}
\pdfoutput=1
\providecommand{\event}{T. Neary, D. Woods, A.K. Seda and N. Murphy (Eds.):\\ The Complexity of Simple Programs 2008.}
\providecommand{\volume}{1}
\providecommand{\anno}{2009}
\providecommand{\firstpage}{130}
\usepackage{breakurl}
\usepackage{amsthm}
\usepackage{graphicx}

\usepackage{amssymb}

\def \cue#1{
Fig.\ref{#1}
\begin{figure}
\hfill
\begin{center}
\includegraphics[width=\textwidth]{#1} 
\end{center}
\hfill
\caption{#1}\label{#1}  
\end{figure}
}

\title{The Pagoda Sequence: a Ramble through Linear Complexity, Number Walls, 
        D0L Sequences, Finite State Automata, and Aperiodic Tilings}

\author{Fred Lunnon
    \email{Fred.Lunnon@may.ie}
    \institute{National University of Ireland Maynooth, Co. Kildare, Ireland}
}
\def\titlerunning{The Pagoda Sequence}
\def\authorrunning{F. Lunnon}

\begin{document}
\maketitle
\begin{abstract}
We review the concept of the number wall as an alternative to the traditional
linear complexity profile (LCP), and sketch the relationship to other topics
such as linear feedback shift-register (LFSR) and context-free Lindenmayer
(D0L) sequences. A remarkable ternary analogue of the Thue-Morse sequence
is introduced having deficiency 2 modulo 3, and this property verified via
the re-interpretation of the number wall as an aperiodic plane tiling.
\end{abstract}

%
%

\section{Introduction}
In the early 1970's the availability of the Berlekamp-Massey algorithm led to
the emergence of the Linear Complexity Profile (LCP), as a measure of how well
a sequence of (say) binary digits could be approximated by a Linear Feedback
Shift-Register (LFSR) --- a topic of some practical importance in the design
of cryptographic key-stream sequences.
\par
A less established alternative, previously known to rational approximation
specialists by the somewhat unimaginative term {\sl C-table}, is the
{\sl number wall} --- an array of Hankel determinants formed from consecutive
intervals of the sequence --- which lends itself better to geometrical
interpretation than the traditional LCP.
\par
An algorithm for number-wall computation, generalising the classical
Jacobi recurrence to the previously intractable case of zero determinants,
was later discovered by the author, who typically then failed to get around
to actually publishing it for another 25 years. It is applicable to sequences
over any integral domain, and with care can be implemented to cost constant
time per entry computed.
\par
A particular area of interest involves sequences whose complexity according
to this model is in some way extreme, such as that proposed by Rueppel with
a so-called `perfect' LCP. Sequences with `perfect' number-walls are
harder to find, in fact over a finite domain they appear not to be possible:
a probabilistic argument gives approximate bounds on the depth of such tables,
confirmed by computer searches modulo 2 and 5.
\par
Despite this in 1997 was discovered a remarkable sequence with modulo 3
deficiency 2, that is its ternary number-wall contains only isolated zeros
--- or in plainer language, no linear recurrence or LFSR of order $m$ spans any
$2m+2$ consecutive terms, for any order at any point. More remarkably still,
computational evidence suggests that the same sequence has deficiency 2 modulo
other primes of the form $p = 4k-1$.
\par
The construction of this {\sl Pagoda} sequence resembles that of the classical
square-free Thue-Morse ternary sequence: an auxiliary sequence is generated via
a D0L system, then mapped to the target sequence via a final extension morphism.
Such D0LEC (D0L with extension and constant width) or `automatic' sequences
have some claim to form a natural complexity class immediately above the LFSR
class, combining greater flexibility with accessible distribution properties.
\par
The proof of the deficiency modulo 3 was finally accomplished two years
later, involving the recasting of the number wall as a tiling of the plane ---
essentially a two-dimensional D0LEC --- by a tesselation using 107 different
varieties of tile. Proof for other primes remains elusive.
\par

\section{Linear Complexity}
A sequence $[S_n]$ is a {\sl linear recurring} or {\sl linear feedback shift
register} (LFSR) sequence of order $r$, when there exists a nonzero vector
$[J_i]$ (the {\sl relation}) of length $r+1$ such that
$$\sum_{i=0}^rJ_iS_{n+i}\ =\ 0 \quad\hbox{for all integers $n$.}$$
If the relation has been established only for $a\le n\le b-r$ we say that
the relation {\sl spans} $S_a,\ldots,S_b$, with $a=-\infty$ and $b=+\infty$
permitted.
\par
Sequences may have as elements members of any integral domain: in applications
the domain will usually be the integers or some prime (often binary) finite
field. LFSR sequences over finite fields are discussed comprehensively in
\cite{Lid97} \S6.1--6.4.
It must be emphasised that the same sequence may have very different
linear complexity behaviour, according to the domain considered:
this caveat will apply in particular to profiles and walls of integer sequences
modulo a prime, often 2 or 3.
\par
Of practical importance in the design of secure cryptographic key-stream
sequences is the question of how well a binary sequence is approximated by
(one or more) LFSR's.
Developed in the early 1970's, the {\sl (Shifted) Linear Complexity Profile}
(LCP/SLCP) represented an attempt to establish a relevant quantitive formalism:
given $[S_n]$, its LCP is an auxiliary sequence with $m$-th term the order of
the minimal LFSR spanning segment $S_0,\ldots,S_{m-1}$;
the SLCP generalises this reluctantly into two dimensions by considering the
order of $S_n,\ldots,S_{n+m-1}$, where both $m$ and $n$ vary.
\par
In recent years linear complexity has made little progress; and it is my
contention that the major culprit is the accidental manner in which LCP's
were contrived. The Berlekamp-Massey algorithm had recently been developed,
providing a means of computing the minimal relation spanning $n$ terms of a
sequence in time quadratic in $n$.
This seems then to have been seized upon by both coding and complexity
communities --- the latter simply discarding the components $[J_i]$ of the
relation, retaining only the order~$r$.
\par
To introduce a personal note at this point, I have to confess to having never
felt comfortable with Berlekamp-Massey: its application is tricky ---
for instance, the intermediate vectors it generates cannot be relied
upon to represent relations spanning a prefix of the segment --- and
its proof (see \cite{Lid97}) strikes me as both complicated and lacking
obvious direction.
\par
A more natural and elementary alternative considers instead the
simultaneous linear equations for the relation components $[J_i]$ in terms of
the sequence elements $[S_n]$. Easily, these have a solution just when the
{\sl Toeplitz} determinant [or with an extra reflection, {\sl Hankel} or
{\sl persymmetric}]
$$S_{mn}\ =\ \left|\matrix{S_n&S_{n+1}&\ldots&S_{n+m}\cr
  S_{n-1}&S_{n}&\ldots&S_{n+m-1}\cr
  \vdots&\vdots&\ddots&\vdots\cr
  S_{n-m}&S_{n-m+1}&\ldots&S_n\cr}\right|$$
vanishes.
\par
A zero entry $S_{mn}$ indicates a relation of order $r\le m$
spanning the segment $[S_{n-m},\ldots,S_{n+m}]$. If the sequence is in fact
generated by a single LFSR of order $r$, the table will be zero from row $r$
onwards: therefore this {\sl number-wall} bears the same relation to an LFSR
sequence as does the difference table to a polynomial sequence (where $S_n$
is a polynomial function of $n$); in fact, one generalises the other, to the
extent that every polynomial sequence of degree $r-1$ is de facto LFSR of
order $r$, with relation given by the vanishing of its $r$-th difference.
\par
These determinants can also be computed in quadratic time, via an algorithm
not only progressive [so the time becomes effectively linear for a table
of many values], but beguilingly simple and symmetrical, and classical ---
being a special case of a well-known pivotal condensation rule or extensional
identity credited variously to Sylvester, Jacobi, Desnanot, Dodgson, Frobenius: 
$$S_{m,n}^2\ =\ S_{m+1,n} S_{m-1,n}\ +\ S_{m,n+1} S_{m,n-1}.$$
\par
Unfortunately, formulating a corresponding recursive algorithm, expressing
each row in terms of the two previous,
\begin{eqnarray*}
  S_{-2,n}\ &=&\ 0,\quad S_{-1,n}\ =\ 1,\quad S_{0,n} = S_n,\cr
  S_{m,n}\ &=&\ \bigl(S_{m-1,n}{}^2\ -\ 
  S_{m-1,n+1}S_{m-1,n-1})\bigr)/S_{m-2,n}\quad\hbox{for $m > 0$}\cr
\end{eqnarray*}
reveals immediately a major flaw: once a zero has been encountered,
computation is unable to proceed beyond the subsequent row, on account of
division by $S_{m-2,n} = 0$.
\par
One of the more elementary properties --- already familiar in the guise of
the {\sl Pad\'e block theorem} (see \cite{Gra72}) to Pad\'e table specialists
[who have incidentally been collectively responsible for a remarkable number 
of bogus proofs of it] is that zero entries occur only as continuous square 
regions, surrounded by an {\sl inner frame} of nonzeros 
[easily seen by the Sylvester identity to comprise a geometric sequence along
each edge].
\par
Some time around 1975, I succeeded in generalising the recursion to bypass such
zero entries. Ironically (given my original motivation) even the statement
of these {\sl frame theorems} demands sufficient preliminary background to
necessitate relegation to appendix A;
and their convoluted and technical proof required several attempts, finally
involving a combination of methods from ring theory, analysis and algebraic
geometry, and sustained over a period of more than a quarter of a century
\cite{Lun01}.
\par
[John Conway, who took an early interest in this topic, christened the zero
regions {\sl windows}, and the table a {\sl wall of numbers}, \cite{Con96}.
Apparently, on first encountering these results, he transcribed them for safe
keeping onto his bathroom wall (the way one does); but having moved house by the time the book came to be written, was obliged to rely on memory, and as a result 
(to his evident embarrassment) committed two separate typographical errors in
restating them.]
\par
There is plainly a close relationship between SLCP's and number walls --- see
\cite{Ste92} for example. However, the more symmetrical definition of the latter
considerably facilitates the deployment of genuinely two-dimensional geometry
in their investigation, as we shall see later; in contrast, the (diagonal,
one-dimensional) generating function technique --- encouraged by the LCP
paradigm --- is for example unable to probe the central diamond of a number wall
at all.
\par

\section{D0L and D0LEC systems: Thue-Morse sequence}
A deterministic context-free Lindenmayer (D0L) system is defined to be
a substitution system where there is only one production for each symbol;
all productions are applied simultaneously; and production is iterated,
starting from some distinguished (stable) symbol, so generating an infinite
sequence.
\par
We further define D0LE to mean extended by a final (single-shot) substitution,
usually to an alphabet distinct from that used by the generation stage;
and D0LEC to mean that both morphisms (sets of production rules) have constant 
width on the right-hand side --- for instance, no sneaky null symbols,
mapping to the empty string! [Much the same idea appears elsewhere under the
umbrella of `automatic sequences' --- see `image, under a coding, of a
$k$-uniform morphism' in sect.~6 of \cite{All03}.] 
\par
Why should D0L (and particularly D0LEC) systems be worthy of study?
LFSR systems arise naturally in a number of applications (signal-processing,
cryptography), and the number wall is a natural tool with which to investigate
them. When we come to study number walls in turn, their extremal behaviour is
observed to occur for D0LEC sequences (which incidentally arise in other
unrelated applications as well). So D0LEC sequences in some sense
constitute a natural third layer in a complexity hierarchy commencing thus:
polynomial sequences, LFSR sequences, D0LEC sequences, $\ldots$
\par
The distribution properties of these sequences can easily be established,
using classical Markov-process methods \cite{Fel57}.
Another bonus is algorithmic: the D0LEC paradigm permits both the computation
of a distant term $S_n$ of a sequence, and furthermore the inversion of this
process to recover $n$ from $S_n$ (where this is single-valued), in time of
order $\log n$ by means of a finite-state automaton --- see \cite{All03}.
\par
In illustration of these ideas, we turn now to consider the Thue-Morse sequence.
[This is conventionally constructed as the fixed point of the morphism
$0 \to 01,\ 1 \to 10$; however, the following indirect construction proves
more illuminating.] Recall that a sequence of symbols is {\sl square-free}
when no factor word (of consecutive symbols) is followed immediately by a copy
of itself; similarly, a sequence may be {\sl cube-free}, {\sl power-free}.
\par
Consider the D0L system on 4-symbols defined by the generating morphism
$$ \Phi: A \to BC,\ B \to BD,\ C \to CA,\ D \to CB;$$
notice the symmetry of $\Phi$ under the permutation $(AD)(BC)$.
Starting from $B$ and applying $\Phi$ repeatedly gives
what turns out to be a square-free right-infinite quaternary sequence:
$$[V_n]\ =\ BDCBCABD\ CABCBDCB\ CABCBDCA\ BDCBCABD\ \ldots$$ 
[This could be made left- and right-infinite by starting with $AB$ or $CB$
and fixing the origin in the centre; but then $\Phi^2$ rather than $\Phi$
would be required to obtain stability.]
\par
The final morphism
$$ A \to 0,\ B \to 0,\ C \to 1,\ D \to 1, $$
now yields the classical cube-free binary Thue-Morse sequence
$$[T_n]\ =\ 01101001\ 10010110\ 10010110\ 01101001\ \ldots,$$
explicitly $T_n$ equals the sum modulo 2 of the digits of $n$ when expressed
in binary. Alternatively, the final morphism
$$ A \to 0,\ B \to 1,\ C \to 2,\ D \to 0, $$
yields the related ternary sequence
$$[U_n]\ =\ 10212010\ 20121021\ 20121020\ 10212012\ \ldots,$$
which can be shown to be square-free. Proofs are given in appendix B;
they bear comparison with rather complicated ad-hoc arguments available
elsewhere, e.g. \cite{Lot83}.
\par
More significantly, other final morphisms may be tailored to produce new
sequences, such as 
$$ A \to 11,\ B \to 01,\ C \to 10,\ D \to 00, $$
yielding a binary sequence which has no squared words of length exceeding 6:
$$01001001\ 10110100\ 10110110\ 01001001\ \ldots,$$
and
$$ A \to 1101,\ B \to 0011,\ C \to 1000,\ D \to 0010, $$
with no squares exceeding length 4 (optimal): 
$$01110010\ 10000111\ 10001101\ 01110010\ 10001101\ 01111000\ \ldots$$
\par

\section{Average versus Extremal Walls}
We propose to illustrate the discussion using an interactive Java application
which displays number walls of various special sequences modulo a given prime.
Entries are encoded as coloured pixels: white for 0, black for 1, grey for 2;
or red for 2, green for 3, blue for 4, etc. interactively; the sequence runs
along two rows from the top edge.
Program source {\sl ScrollWall.java} is available from the author; 
\par
The implementation is based on the frame theorems (appendix A), incorporating
an enhancement to obviate searching when circumnavigating a large window.
[The binary case is particularly simple, to the extent that an
exceptionally efficient implementation is feasible in the form of a 44-state
cellular automaton based on the Firing-Squad Synchronisation Problem (FSSP)
--- see \cite{Min67}; \cite{Lun01} sect.7.]
The given finite segment must be extended into a periodic sequence, to avoid
algorithmic complications resulting from the presence of a boundary: therefore
in general, only the triangular north quarter of a (square) graphical display
is significant; although in special cases, intelligent choice of segment
length $n$ may improve this situation. Since a sequence with period $r$ is LFSR
with order at most $r$, the number of nonzero rows (including the initial row of
empty determinants) for a segment of length $n$ columns must be at most $n+1$.
\par
So as to have something for later comparison, we first take a look at
a `typical' number wall. When the domain is a finite field with $q$ elements,
it can be shown that for a random sequence, the asymptotic mean density of
size-$g$ (or $g\times g$) windows exists (in some suitably weak sense),
and equals $$(q-1)/(q+1)q\ \cdot\ 1/q^g ;$$ 
for example, \cue{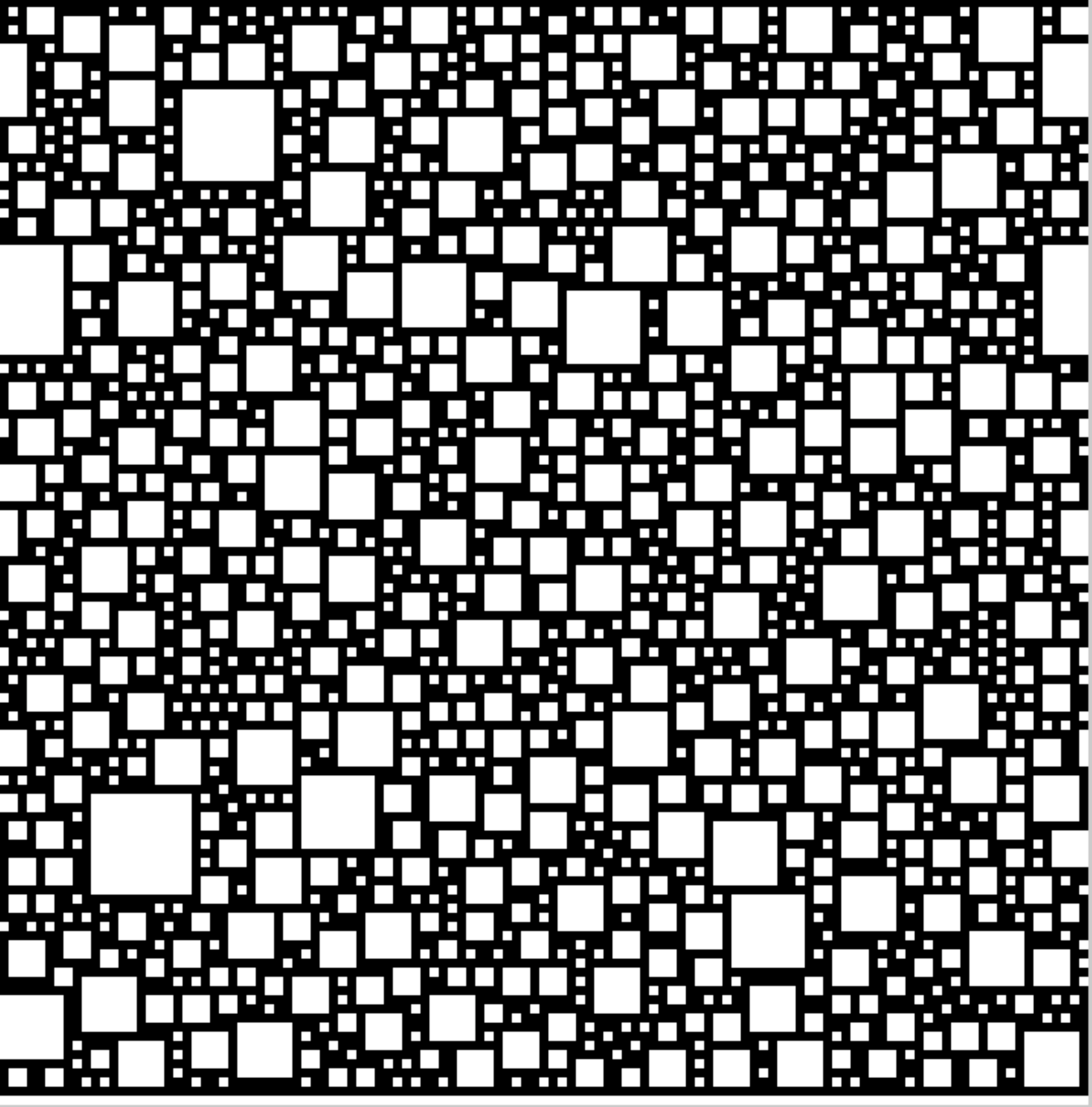}, \cue{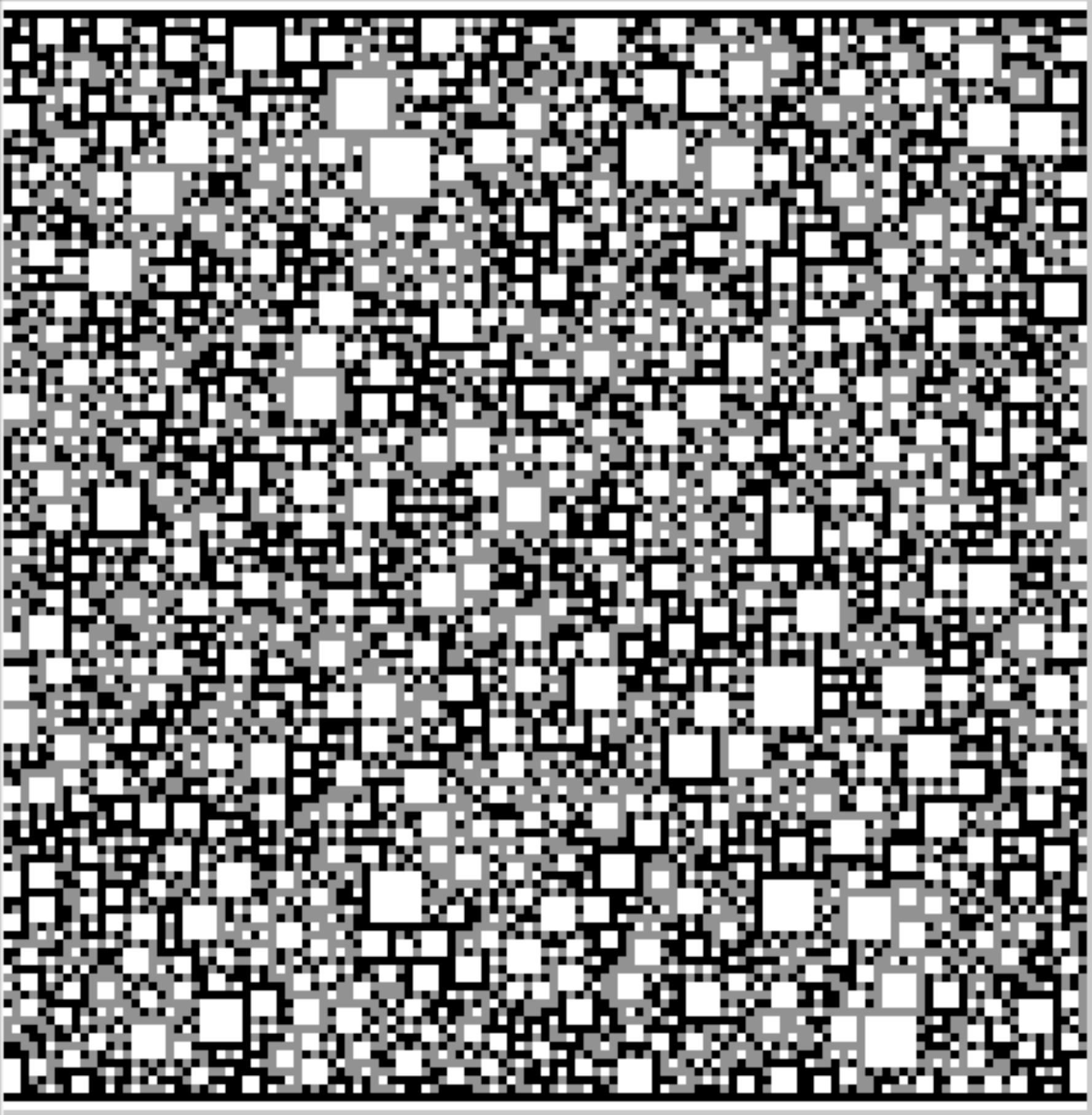} modulo $q = 2,3.$
\par
It is tempting to employ this result as a test for randomness: for example,
counting the numbers of windows of each size in a suitably large portion of
the wall, then applying the $\chi^2$ test to the frequencies.
A discouraging counterexample is the sum of the Thue-Morse and Rook sequences
modulo 2\linebreak (see \cue{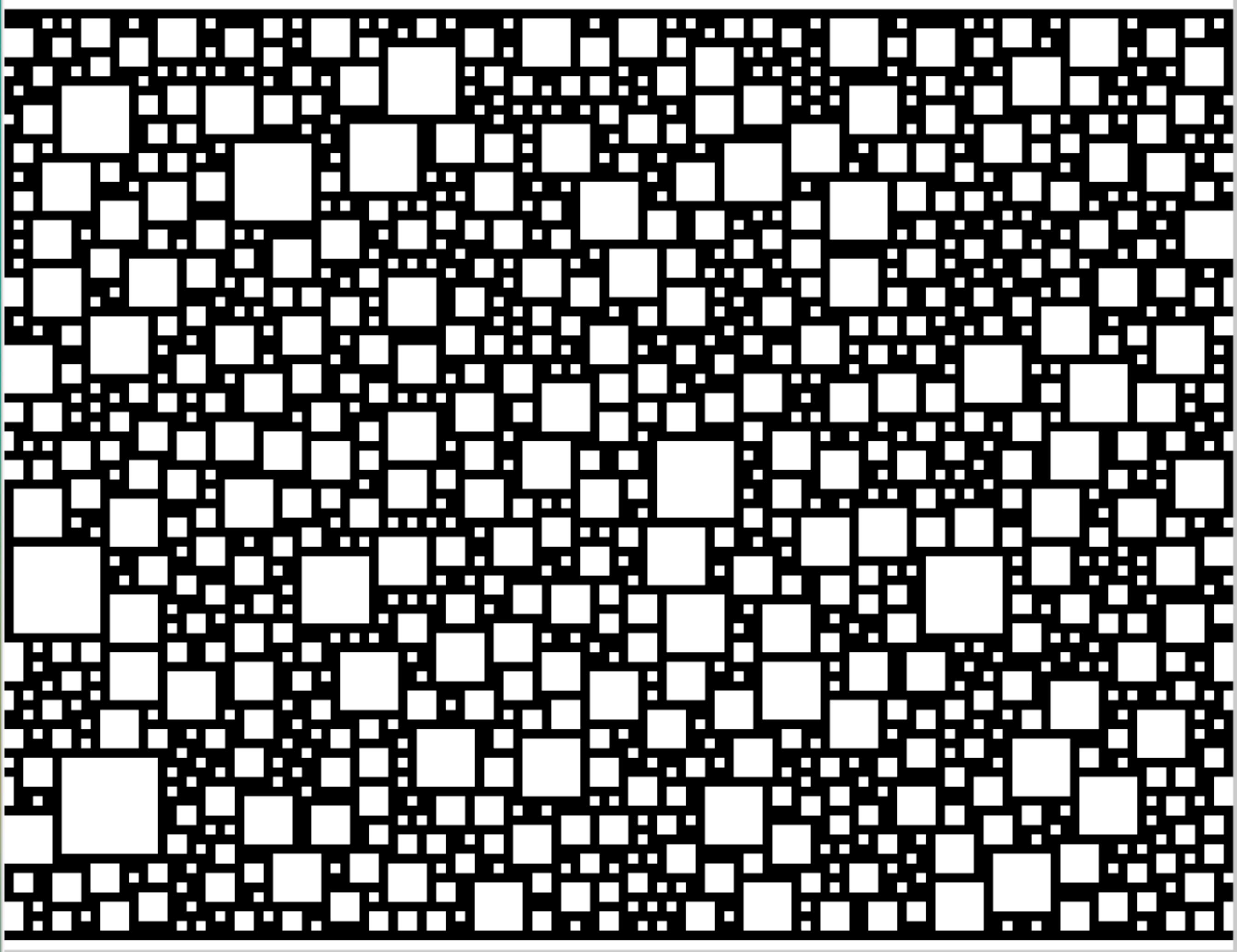}), which passes this test with flying colours,
despite being generated by the 8-symbol D0LEC system: 
$$A\to Ab, B\to Ad, C\to Cb, D\to Cd, a\to aB, b\to aD, c\to cB, d\to cD;$$ 
$$A\to 0, B\to 0, C\to 1, D\to 1, a\to 1, b\to 1, c\to 0, d\to 0.$$ 
\par
Our major target in this essay is the investigation of extremal walls: by which
is meant, the extent to which a number wall may deviate from typical window
distribution.
One pretext for this activity is exposure of the limitations of the paradigm;
but it might be more honest to prefer the serendipitous justification,
that some of the graphic art so produced is simply rather striking [and might
become more so, were the author's casually primitive palette to be refined!]
\par
The (not overly impressive) example of the Rueppel sequence makes a point about
the limitations of the original LCP concept.
Its definition is
$$ S_n = \cases{
1 &if $n = 2^k - 1$ for some $k$;\cr 
0 &otherwise.\cr} $$
It was proposed as an example of a binary sequence having `perfect' LCP, which
in number-wall terms implies a continuous nonzero diagonal staggering from one
corner to the opposite. Elsewhere though, its wall is perfectly appalling,
composed almost exclusively of windows increasing exponentially in size
(see \cue{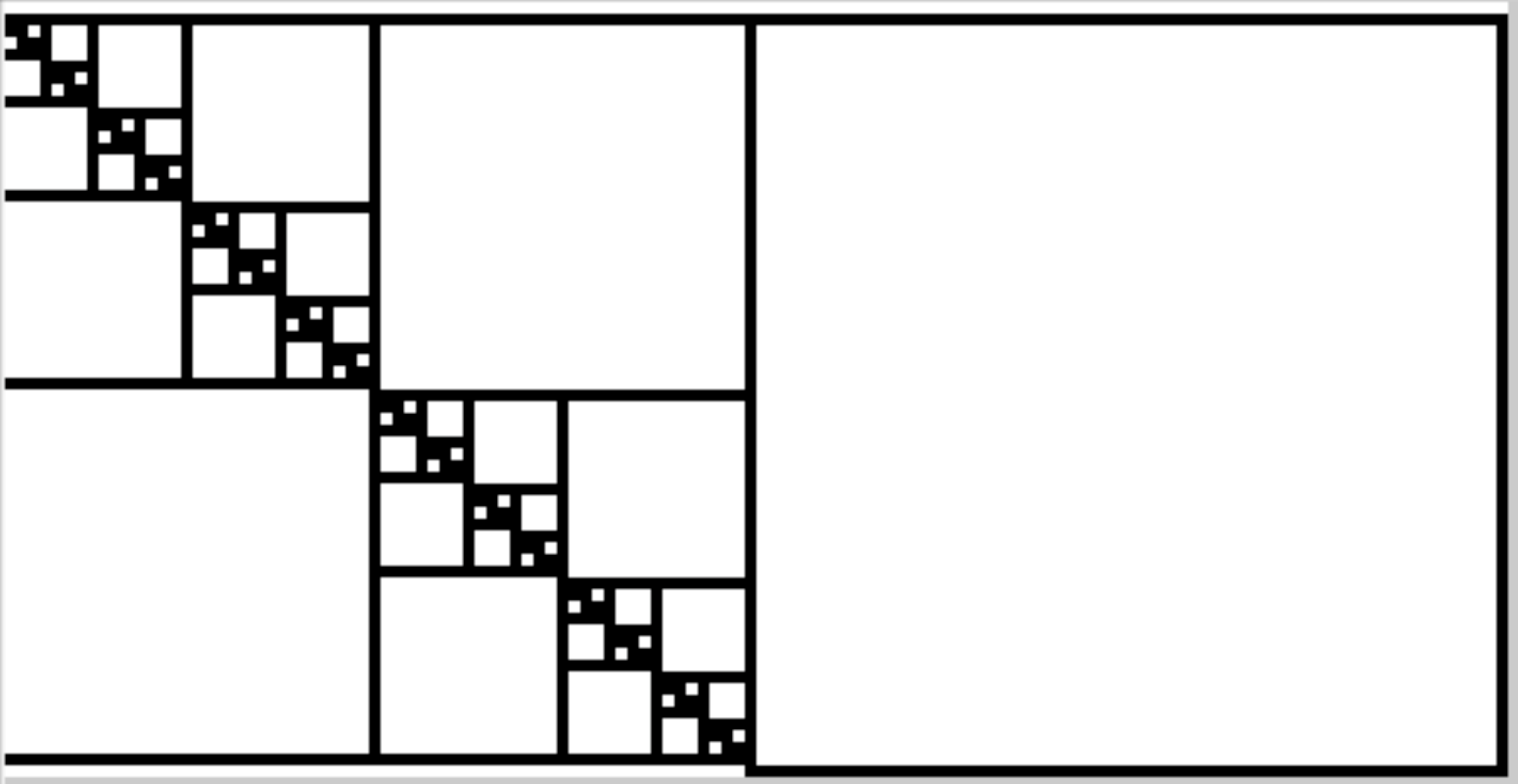}).
\par
But it suggests an analogous though considerably tougher challenge, which we
proceed to take up: to determine the extent to which a (binary, say) wall can
avoid zero entries. A combinatorial argument based on the frame theorems shows
easily that any extended region of the wall has local zero-density at least
$1/5$ --- the minimal pattern has isolated zero entries occurring a knight's
move apart.
Globally, this minimal pattern can occupy at best an infinite central diamond,
the rest of the wall comprising a fractal-like pattern of increasing windows and
finite minimal diamonds, see \cue{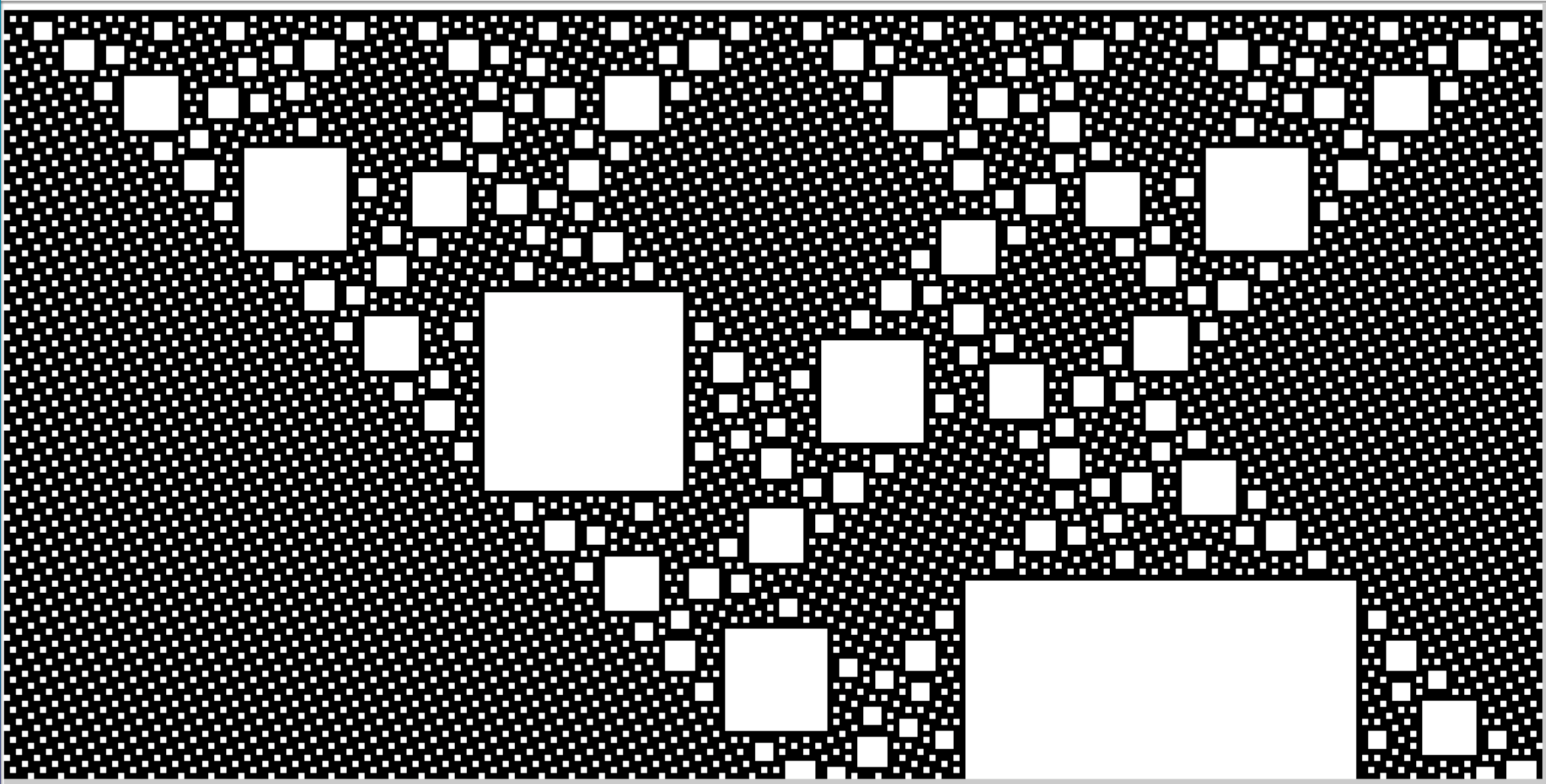}.
\par
To explicitly construct the sequence with this wall, first define the {\sl Rook}
sequence $[R_n]$ as the digit preceding the least-significant $1$ in the binary
expansion of $n$, or $0$ if $n=0$ --- compare with Thue-Morse.
E.g. if $n = 104 = 1101000$
in binary, the final $1$ is 3 digits along from the end, and the 4-th digit
along is $R_{104} = 0$. $[R_n]$ is a binary sequence, and a recursion for it is
$$ R_{-n}\ =\ 1-R_n\hbox{ for }n \ne 0;\quad R_{2n}\ =\ R_n;
\quad R_{2n+1}\ =\ n\bmod 2.$$
The first few values for $n\ge0$ are
$$[R_n]\ =\ 00010011\ 00011011\ 00010011\ 10011011\ \ldots $$
Finally, define the {\sl Knight} sequence by
$$K_n = R_{n+1} - R_{n-1} \pmod 2 .$$
\par
For ternary walls, the situation is rather similar: an essentially unique
nonzero local pattern exists, composed of alternating zigzag stripes of $+1$
and $-1$ resp. Globally, this motif can be replicated only within a central
diamond; the remainder of the wall is now rather sparse, not unlike
the Rueppel wall, see \cue{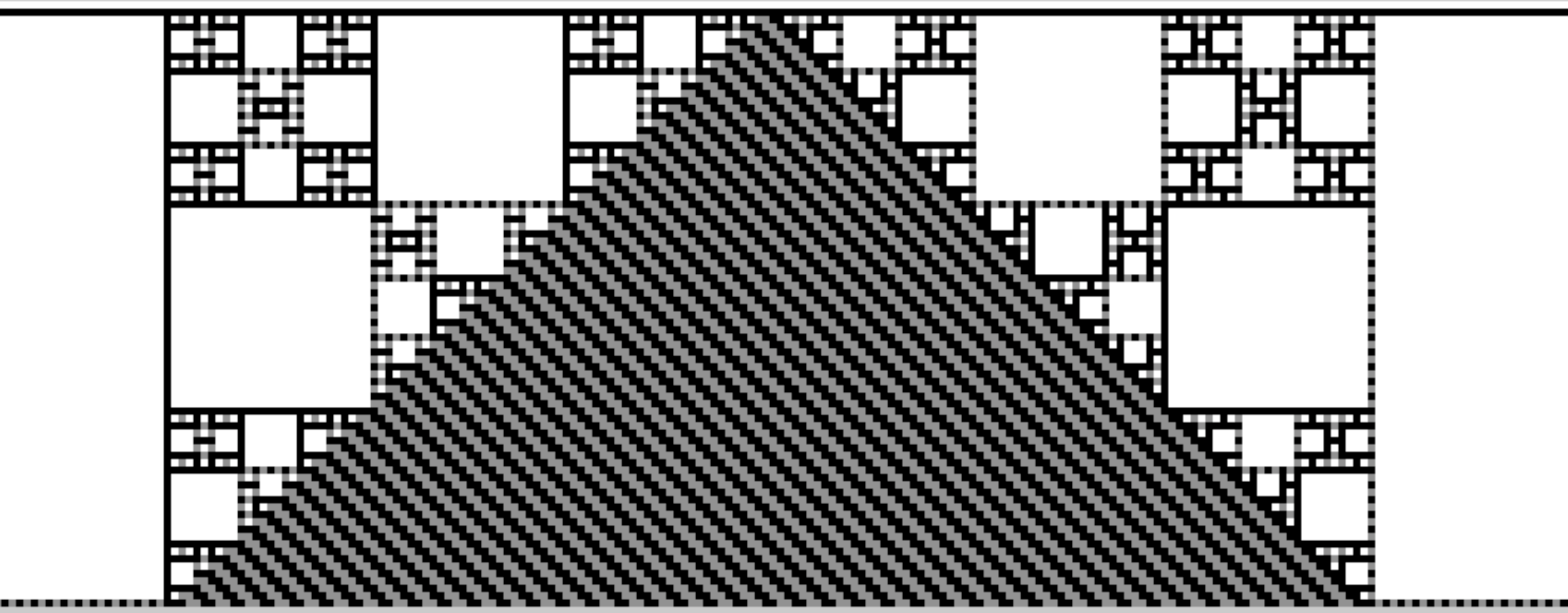}.
An explicit expression for this sequence is clumsy,
but it has the D0LEC definition:
$$ A \to ACB,\ B \to BCB,\ C \to EDF,\ D \to DDD,\ E \to EDD,\ F \to DDF; $$
$$ A \to 1,\ B \to 0,\ C \to 1,\ D \to 0, E \to 2,\ F \to 2. $$
Starting from $A$, the first few terms of generated and final sequences are 
$$ ACBEDFBCB\ EDDDDDDDF\ BCBEDFBCB\ \ldots; $$
$$ [Z_n]\ =\ 110202010\ 200000002\ 010202010\ 200000000\ 000000000\ \ldots$$
\par
Accepting that a total absence of zeros (on rows $m\ge-1$) is not possible,
we can instead attempt in various ways to circumscribe their occurrence.
Rather than become involved in somewhat recondite questions regarding what
exactly might be meant by the term {\sl density} in this context, we shall
consider the more concrete problem of bounding the size of the windows.
\par
A simple probabilistic argument can be mounted suggesting that, when the domain
is a finite field with $q$ elements, the size of the maximum window occurring
within the first $m$ rows of a wall will be of the order of $\log_q m$;
and more strongly, that the probability of a sequence having no windows larger
than this bound is zero. [This contrasts with the situation for square-free
sequences, where the corresponding probability is nonzero for $q > 2$.]
\par
With this in mind, we conducted a search for binary sequences with the greatest
number of rows having no window of size $d$ or greater, for small values of the
[in LCP jargon] {\sl deficiency} $d$. This endeavour is highly speculative:
first the critical depth $m$ must be established such that no satisfactory
sequence exists with greater depth; then a sufficiently long segment constructed
for an evident period to become established.
\par
The resulting handful of sequences is shown in the table:
all are periodic with period $t$, and the order $r$ equals the final depth $m$
satisfying the deficiency bound --- that is, as soon as the bound fails,
the entire wall vanishes --- and $m$ seems to increase exponentially with $d$
as expected. Confidence in these results is encouraged by the presence of
adventitious symmetries, such as the 0-1 alternating subsequence at odd
positions of case $d = 4$. See \cue{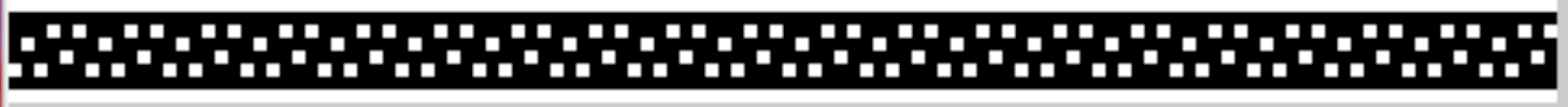},\cue{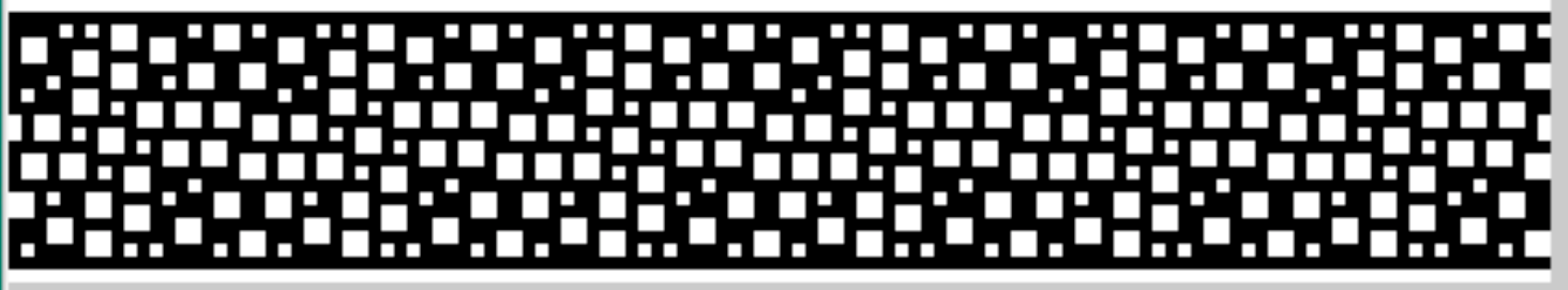},\cue{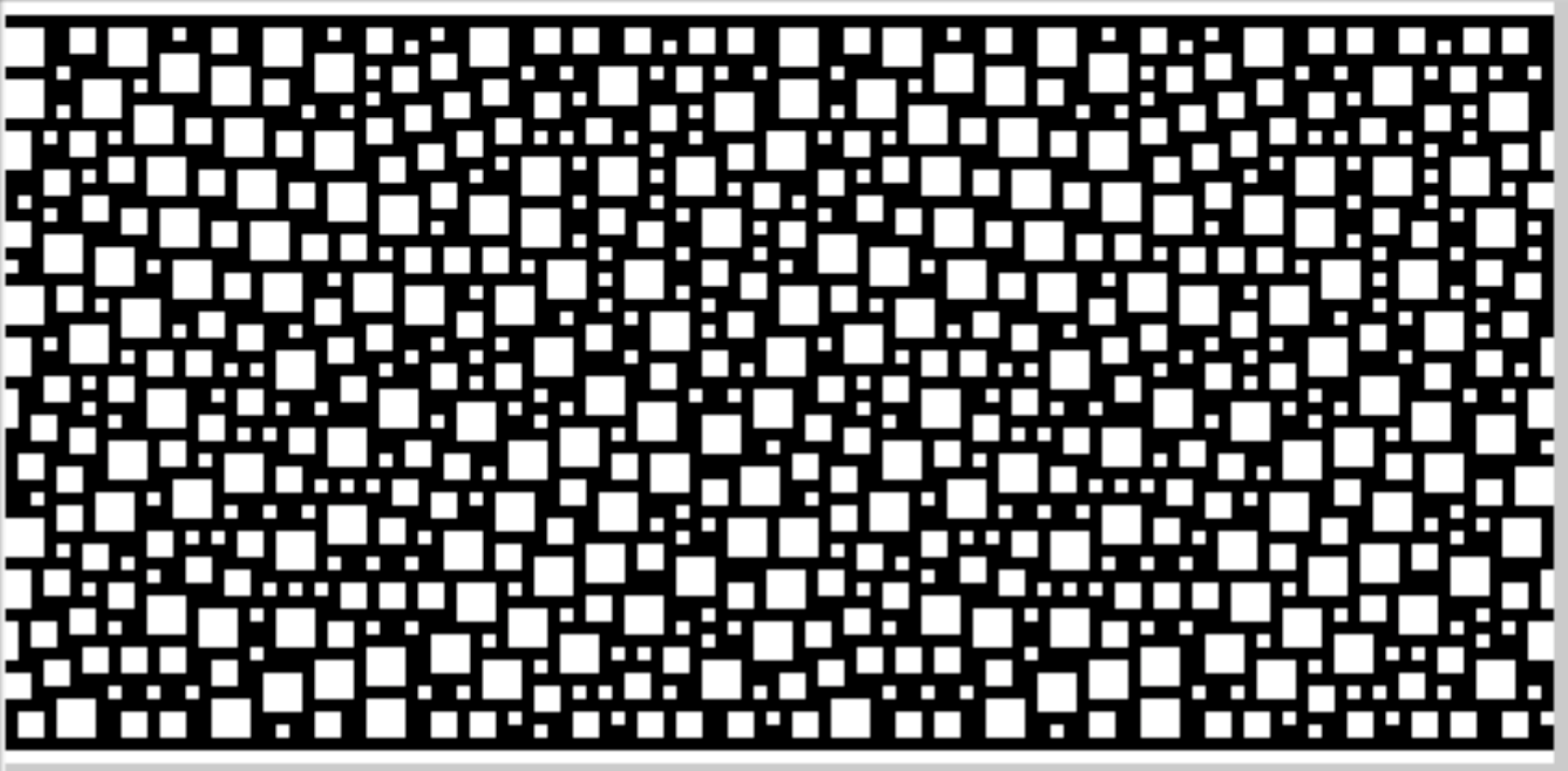}.
\begin{table}
\begin{tabular*}{0.9\textwidth}{@{\extracolsep{\fill}} | r | r | r | r | | l | }
  \hline
  $d$ & $m$ & $r$ & $t$ & period \cr
  \hline 
   1 &  1  &  1 & 1  & [1] \cr
  \hline
   2 &  5  &  5 & 6  & [111010] \cr
  \hline
   3 & 19  & 19 & 20 & [1111010100\ 1111010010] \cr
  \hline
   4 & 56  & 56 & 60 & [0001100100\ 0110110011\ 0001101100 \cr
     &     &    &    & \quad 1110110001\ 1001001100\ 1110010011] \cr
  \hline
   5 & 95+ &  ? &  ? & (none detected in 800 terms) \cr
  \hline
\end{tabular*}
\end{table}

Now what about ternary walls? Deficiency $d = 1$ is disposed of trivially,
by the period-4 sequence [1122\ldots] with $m = r = 2$. But when our search
program is let loose on $d = 2$, the first of a number of strange things
happens --- or in this case, fails to happen --- the depth goes on increasing
indefinitely, while (necessarily) no period ever properly quite stabilises.
To cut quite a long story short, the object which eventually emerges is
a remarkably simple D0LEC, has deficiency-2 to any depth we care to examine,
and turns out to be essentially identical to the Knight $[K_n]$ --- seen
earlier in an unrelated context!
\par
To be precise, with $R_n$ the Rook sequence as above, the {\sl Pagoda} sequence
is defined by
$$P_n = R_{n+1} - R_{n-1} \pmod 3 .$$
The ternary number wall is shown at \cue{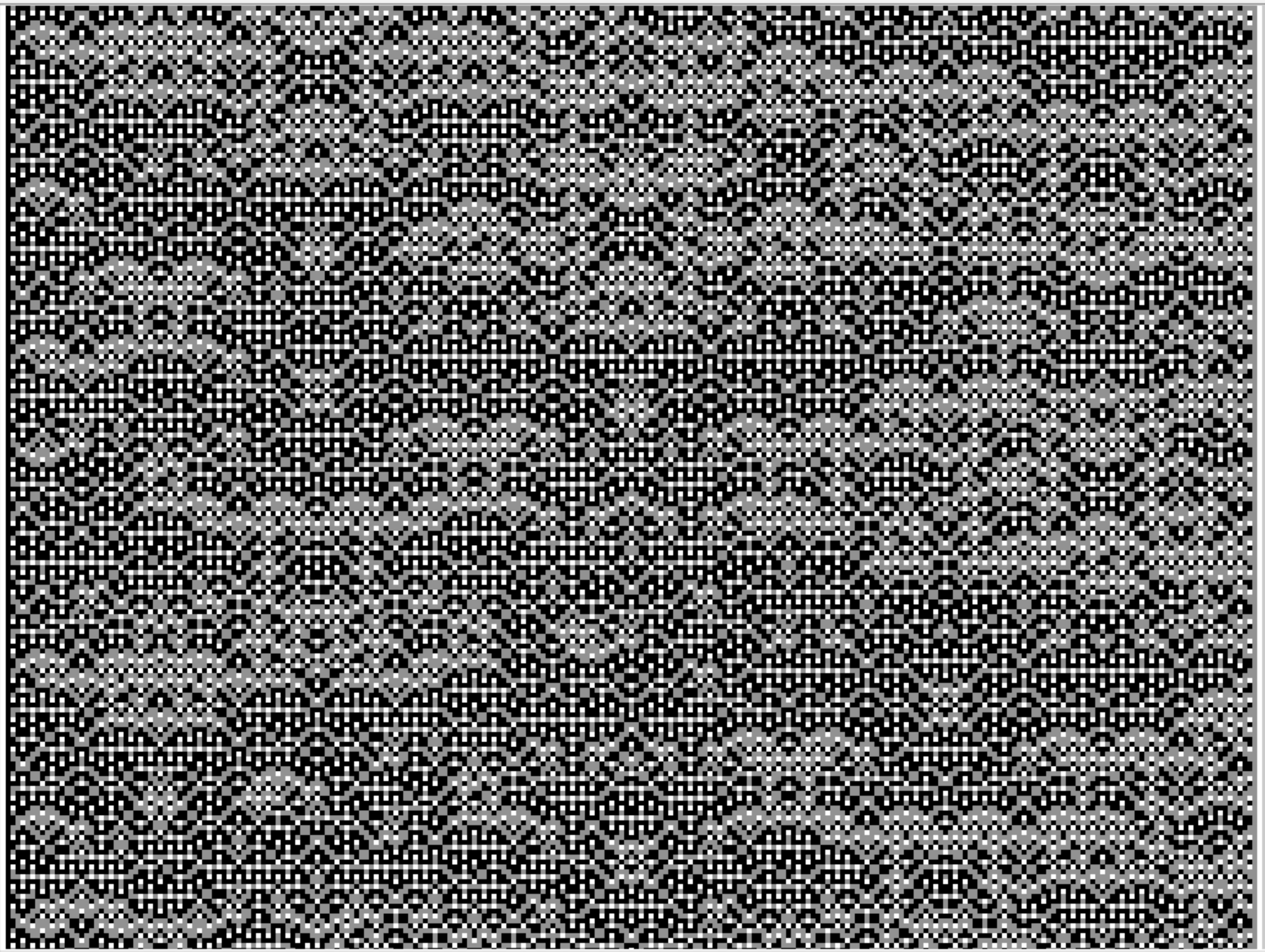}; the symmetrical, fractal-like
filigree structures for which it was christened are more easily appreciated
after rotation through a quarter-turn, the sequence running down the left side.
\par
Examination of substantial portions of the number-walls of this sequence modulo
$$p = 3,7,11,19,23,31,43,47,59,67,71,79$$
encourages the conjecture that its deficiency remains equal to 2 modulo
{\sl any} prime $p = 4k-1$; modulo $p = 83$ however, this elegant simplicity is
confounded by the presence of numerous windows of size 2, together with what
appears to be a splendidly lone specimen of size 3 commencing at\linebreak entry
$m = 105, n = 188$ [a specimen discovered only during protracted investigation
of an apparent compiler bug causing ScrollWall to report spurious runtime
errors].
\par
The Pagoda was not the first, nor the last of its kind to be discovered:
but all these, along with the Knight and Rook sequences, are closely
intertwined, in a manner notably reminiscent of our earlier analysis of
the Thue-Morse family. Consider the 4-symbol D0L system
$$ A \to AB,\ B \to AD,\ C \to CB,\ D \to CD; $$
applied to $A$ this generates 
$$ [V_n] = ABADABCD\ ABADCBCD\ ABADABCD\ CBADCBCD\ \ldots$$
[which can be made infinite both ways by starting instead from $DA$ and choosing
the origin to be the first symbol of the (inflated) original $A$.]
Applying the final morphism
$$ A \to 2201,\ B \to 0211,\ C \to 0221,\ D \to 1201 $$
yields the Pagoda sequence $[P_n]$, for $n \ge 0$ [or all $n$];
$$ A \to 1101,\ B \to 0111,\ C \to 0111,\ D \to 1101 $$ 
yields the Knight sequence $[K_n]$.
\par
Other deficiency-2 variations on the Pagoda may be concocted by varying the
final morphism.
Also notice that the generator is not symmetric under either transposition
$(AC)$ or $(BD)$: so these provide a set of 4 distinct generators, each of
which could be used to yield an alternative quaternary sequence.
Applying the final morphism
$$ A \to 0,\ B \to 0,\ C \to 1,\ D \to 1 $$
to any of these alternatives yields the same binary sequence, the Rook $[R_n]$.
\par
Modulo these variations, and the continuum of variants obtained by shifting
the origin repeatedly during generation, it seems quite plausible that the
Pagoda is the unique ternary sequence with this deficiency. 

\section{Pagoda Tiling Proof}
At this point, we have some probabilistic arguments and experimental evidence
to support the conjectures that:
\begin{itemize}
\item[] If $p \bmod 4 = 1$ or $p = 2$, then the maximum depth $m$ to which
deficiency $d$ can be maintained by any number wall modulo $p$ is finite,
bounded by order $\log_p d$;
\item[] If $p \bmod 4 = -1$, then the number wall modulo $p$ of the
Pagoda sequence has bounded deficiency (dependent only on $p$) to any depth;
in particular, for $p = 3$ we have $d = 2$ (only isolated zeros).
\end{itemize} 
\par
To actually prove any of these claims poses a considerable challenge.
A conventional approach to the Pagoda conjecture might involve explicit
algebraic evaluation of the Toeplitz determinants, modulo 3, modulo $p$,
or over the integers: while there is some numerical structure visible here 
which might form a basis for an inductive construction, overall this prospect
is not promising. 
\par
A more unexpected route proves at least partially successful: invoking a
two-dimensional geometrical version of the D0LEC paradigm, extending the
representation of the sequence via $[V_n]$ above, into one of
the entire wall as a plane quasi-crystallographic tiling.
In part this is suggested by close visual inspection of the diagram, which
reveals (at the cost of substantial hazard to eyesight) that the `pagodas'
recurring at various scales throughout the wall are embedded in repetitive
{\sl diamonds}, square regions rotated through a one-eighth turn.
\par
Factors to be taken into account in the formalisation of this concept include:
\begin{itemize}
\item[] Interaction between faces, edges and vertices of tiles;
\item[] Non-trivial point symmetries tiles may possess;
\item[] Choice of an appropriate translation of tiling origin;
\item[] Extent to which tiles are open or closed subsets of the plane;
\item[] Determination of tile size, or D0L extension width; 
\item[] Determination of number of distinct tiles, or D0L symbol count.
\end{itemize}
All these factors, along with other details relevant to implementation only at
a detailed level, need be taken into account in the design of a program to
(as it were) tile a wall --- to specify the precise spatial `inflation'
morphism generating it, along with the extension `pattern' on each tile.
\par
It is natural to align the vertices of a tile with entries of the number wall,
so that an entry at a vertex is shared between 4 adjacent tiles, at an edge
between 2. This presents a conflict between notational clarity and computational
simplicity, resolved by including the entire boundary in tile morphism diagrams
(appendix C); while to actually apply a morphism, the boundary must be shrunk
and displaced by a half-unit along each axis, so that a tile comprises only
complete entries.
\par
In order to verify the frame relations between wall entries, as well as to
keep track of inflation of vertices and edges along with faces of tiles,
the search program actually operates a 4-fold covering of the plane by
overlapping {\sl supertiles} having twice the diameter of the faces.
A post-processor extracts the individual inflations of faces etc,
possibly resulting in tile extents becoming reducible to smaller diameter.
At this stage also, point-group symmetries of `fixed' tiles are extracted;
the number of `free' tiles remaining is then substantially reduced.
\par
For the ternary Pagoda, the program successfully finds a tiling
comprising: 
\begin{itemize} 
\item[] Generator inflation diameter 2 (4 subtiles per inflation);
\item[] Point symmetry group of order 16;
\item[] Tiling origin at $S_{-2,0}$;
\item[] Extent diameter of face 4 (partially spanning 25 wall entries);
\item[] Fixed face count 107, reducing to 13 free;
\item[] Every free face occurring within distance 35 from the origin;
\item[] Free vertex count 39, all within distance 165;
\end{itemize}
The full morphism will be found in appendix C.
Point symmetries comprise products of vertical reflection, horizontal
reflection, complementation of odd rows, complementation of odd columns.
\par
Apart from two restricted to meeting the upper zero half-plane
$m\le -2$, every tile has only isolated zeros: this completes the proof
that the deficiency of the ternary Pagoda equals 2. \qed
\par
But of course, the existence of this tiling permits us to investigate the wall
in much greater detail. For instance, by selectively expanding the D0LEC,
any given entry $S_{mn}$ can now be computed in time of order logarithmic in
the distance $|m| + |n|$ from the origin.
\par
Again, the deficiency theorem may be considerably sharpened:
\begin{itemize}
\item[] If $S_{mn} = 0$ in the ternary wall of the Pagoda sequence $S_n = P_n$,
then the power of 2 dividing $m+2$ exceeds that dividing $n$.
\end{itemize}
In particular, no zeros can occur on rows with $m$ odd, nor on column $n = 0$
(for $m \ge -1$, that is).
\par
Again, applying Markov process analysis to the D0LEC, a $13\times13$ matrix
eigenvalue computation establishes that
\begin{itemize}
\item[] Zero entries in this wall possess asymptotic density in a strong sense,
and this density equals $3/20$.
\end{itemize}
\par
While the tiling method has successfully been applied in other simple cases,
such as the Knight (6 free faces), Rook ($\le 28$), and Thue-Morse,
it has not so far succeeded in tiling the Pagoda modulo 7. Neither is it
known whether or not the number wall of every D0LEC sequence can be so tiled:
a noteworthy test-case in this respect is the `quasi-random' binary Thue-Rook
sum $S_n = T_n + R_n \pmod 2$ mentioned earlier, with window size bounded
apparently by order $(\log m)$.
\par


\appendix

\section{Statement of the Frame Theorems.}
A zero entry $S_{m,n} = 0$ in a number-wall can occur only within a
{\sl window}, that is a square $g\times g$ zero region surrounded by a nonzero
inner frame. The nullity of (the matrix corresponding to)
a zero entry equals its distance $h$ from the (nearest) inner frame edge.
\par
The adjacent diagram illustrates a typical window, together with notation
employed subsequently:
$$\matrix{
&       &E_0       &E_1    &E_2    &\ldots &E_k    &\ldots&E_g   &E_{g+1}   &       &\cr
&F_0    &B,A_0     &A_1    &A_2    &\ldots &A_k    &\ldots&A_g   &A,C_{g+1} &G_{g+1}&\cr
&F_1    &B_1       &{\bf 0}&{\bf 0}&\ldots &{\bf 0}&\ldots&{\bf 0}&C_g       &G_g    &\cr
&F_2    &B_2       &{\bf 0}&\ddots &(P)    &\rightarrow  &      &\vdots
&\vdots    &\vdots &\cr
&\vdots &\vdots    &\vdots &(Q)    &\ddots&       &\uparrow     &{\bf 0}&C_k       &G_k    &\cr
&F_k    &B_k       &{\bf 0}&\downarrow &      &\ddots&(R)   &\vdots&\vdots    &\vdots &\cr
&\vdots &\vdots    &\vdots &       &\leftarrow    &(T)   &\ddots&{\bf 0}&C_2       &G_2    &\cr
&F_g    &B_g       &{\bf 0}&\ldots &{\bf 0}&\ldots&{\bf 0}&{\bf 0}&C_1       &G_1    &\cr
&F_{g+1}&B,D_{g+1} &D_g    &\ldots &D_k    &\ldots&D_2   &D_1   &D,C_0     &G_0    &\cr
&       &H_{g+1}   &H_g    &\ldots &H_k    &\ldots&H_2   &H_1   &H_0       &       &\cr
}$$
\par
The inner frame of a $g\times g$ window comprises four geometric sequences,
along North, West, East, South edges, with ratios $P,Q,R,T$ resp.,
and origins at the NW and SE corners. The ratios satisfy
$$PT/QR\ =\ (-)^g;$$
and the corresponding inner frame sequences $A_k,B_k,C_k,D_k$ satisfy
$$A_kD_k/B_kC_k\ =\ (-)^{gk} \quad\hbox{for $0\le k \le g+1$}.$$
\par
The outer frame sequences $E_k,F_k,G_k,H_k$ lie immediately outside the
corresponding inner, and are aligned with them. They satisfy the relation:
For $g\ge 0$, $0\le k\le g+1$,
$$QE_k/A_k\ +\ (-)^k PF_k/B_k\ =\ RH_k/D_k\ +\ (-)^k TG_k/C_k.$$
\par
Proofs are expounded in \cite{Lun01} sect. 3--4.

\section{Proofs that $[V_n]$, $[T_n]$, $[U_n]$ are power-free.}
We sketch the proofs that these sequences are power-free as claimed.
Suppose that $[V_n]$ is not square-free, and let the earliest
occurrence of its shortest non-empty square start at $V_n$ for $n\ge 0$,
with length $2l>0$.
Suppose $l$ is even: if $n$ is odd, by inspection of $\Phi$ there is only one
possible value for $V_{n-1} = V_{n+l-1}$ given $V_{n} = V_{n+l}$, so there is an
equally short square earlier; if $n$ is even, we can apply $\Phi^{-1}$ to
produce a shorter square of length $l/2$. Suppose on the other hand $l$ is odd:
then for each $i$ one of $V_{n+i}$ and $V_{n+l+i}$ has an even subscript, so
by inspection has to be $B$ or $C$. No new pairs are generated after
$\Phi^3B$, so all words of length 4 occur within $\Phi^4B$; the longest composed
of $B$ and $C$ only is seen to have length 3. So $2l \le 3$, and the square
must be $BB$ or $CC$, which do not occur in $\Phi^3B$. By contradiction,
$[V_n]$ is square-free. \qed 
\par
The inverse morphism from $[U_n]$ to $[V_{n-1}]$ is uniquely defined for
$n \ge 1$, given either of the symbols $U_{n\pm 1}$ adjacent to
$U_n$: it is described by the schema
$$ (2)0(1)\to A,\quad 1\to B,\quad
2\to C,\quad (1)0(2)\to D,$$
where $U_{n\pm 1}$ is parenthesised.
If $[U_n]$ had a square factor with $l > 2$, its inverse image would also be a
square in $[V_n]$, since $A$ and $D$ in corresponding positions necessarily have
an adjacent $B$ and $C$; but $[V_n]$ is square-free. If $l = 2$ the inverse
image might be $AD$ or $DA$, but neither occurs in $[V_n]$. \qed 
\par
The inverse morphism from $[T_n]$ to $[V_n]$ is uniquely defined for
$n \ge 2$, given $T_{n+1}$; it is described by the schema
$$ 0(0)\to A,\quad 0(1)\to B,\quad
1(0)\to C,\quad 1(1)\to D,$$
where $T_{n+1}$ is parenthesised.
If $[T_n]$ had a cubic factor, its inverse image would also be a cube in
$[V_n]$, except possibly for the final symbol; but $[V_n]$ is square-free. \qed 
\par

\section{Pagoda Tiling Morphisms}
Free tiles are numbered 1--13. The `gene' field diagrams the $2\times2$
diamond into which the tile inflates under the generator morphism, each entry
comprising a tile number followed by a combined transformation code.
The `extn' field diagrams the $4\times4$ ternary number-wall diamond into
which the tile finally extends, including boundary shared with neighbouring
tiles. The `symm' field notes all transformations which are symmetries of the
tile. Transformation encoding is as follows: 
\par
\hskip90pt 
\begin{tabular*}{0.5\textwidth}{@{\extracolsep{\fill}} | c | l | }
  \hline
  {\sl code} & {\sl transform} \cr
  \hline 
  \hline
   A & identity \cr
  \hline
   B & reflection along rows \cr
  \hline
   C & reflection along cols \cr
  \hline
   D & half-turn rotation \cr
  \hline
  \hline
   I & identity \cr
  \hline
   J & complement odd rows \cr
  \hline
   K & complement odd cols \cr
  \hline
   L & complement odd rows \& cols \cr
  \hline
\end{tabular*}
\par

\begin{verbatim}
                                       0 
                   2                 0 0 0 
Tile  1: gene  1B      1   , extn  0 0 0 0 0,  symm AI,BI; 
                   4                 1 1 1 
                                       1 
\end{verbatim}
\begin{verbatim}
                                       0 
                   2                 0 0 0 
Tile  2: gene  2       2   , extn  0 0 0 0 0,  symm full; 
                   2                 0 0 0 
                                       0 
\end{verbatim}
\begin{verbatim}
                                       0 
                   3                 1 1 1 
Tile  3: gene  5       7   , extn  1 2 2 0 1;  symm AI; 
                   6                 2 1 1 
                                       1 
\end{verbatim}
\begin{verbatim}
                                       0 
                   3B                1 1 1 
Tile  4: gene  5D      7BJ , extn  1 1 2 0 1;  symm AI; 
                   8                 2 1 1 
                                       1 
\end{verbatim}
\begin{verbatim}
                                       1 
                   10                1 1 2 
Tile  5: gene  9       9BK , extn  1 0 2 0 1,  symm AI,BK; 
                   11                1 1 2 
                                       1 
\end{verbatim}
\begin{verbatim}
                                       1 
                   6C                2 1 1 
Tile  6: gene  7BK     7   , extn  1 0 2 0 1,  symm AI,BK; 
                   10BJ              2 1 1 
                                       1 
\end{verbatim}
\begin{verbatim}
                                       1 
                   3CJ               1 1 1 
Tile  7: gene  12      12BL, extn  1 2 0 1 1,  symm AI,CI; 
                   3J                1 1 1 
                                       1 
\end{verbatim}
\begin{verbatim}
                                       1 
                   6D                1 1 2 
Tile  8: gene  5D      5D  , extn  1 1 2 2 1,  symm AI,BK; 
                   11C               2 2 1 
                                       1 
\end{verbatim}
\begin{verbatim}
                                       1 
                   4D                1 1 1 
Tile  9: gene  13      13BL, extn  1 1 0 2 1,  symm AI,CI; 
                   4B                1 1 1 
                                       1 
\end{verbatim}
\pagebreak 
\begin{verbatim}
                                       1 
                   8C                2 1 1 
Tile 10: gene  5B      5B  , extn  1 1 2 2 1,  symm AI,BK; 
                   10B               1 2 2 
                                       1 
\end{verbatim}
\begin{verbatim}
                                       1 
                   11J               1 1 2 
Tile 11: gene  7J      7BL , extn  1 0 2 0 1,  symm AI,BK; 
                   8B                1 1 2 
                                       1 
\end{verbatim}
\begin{verbatim}
                                       1 
                   8BJ               1 1 2 
Tile 12: gene  9BJ     12  , extn  1 0 2 1 0,  symm AI,CI; 
                   8DJ               1 1 2 
                                       1 
\end{verbatim}
\begin{verbatim}
                                       1 
                   6J                2 1 1 
Tile 13: gene  9BL     12J , extn  1 0 2 1 0,  symm AI,CI; 
                   6CJ               2 1 1 
                                       1 
\end{verbatim}

\bibliographystyle{eptcs}

\begin{thebibliography}{}
\providecommand{\bibitemstart}[1]{\bibitem{#1}}
\providecommand{\bibitemend}{}
\providecommand{\bibliographystart}{}
\providecommand{\bibliographyend}{}
\providecommand{\url}[1]{\texttt{#1}}
\providecommand{\urlprefix}{Available at }
\providecommand{\bibinfo}[2]{#2}
\bibliographystart

\bibliographyend
\end{thebibliography}


\begin{thebibliography}{00}




\bibitem{All03} Allouche, Jean-Paul \& Shallit, Jeffrey \sl Automatic Sequences
\rm Cambridge (2003).
\bibitem{Con96} Conway,~J.~H. \& Guy,~R.~K. \sl The Book of Numbers
\rm Springer (1996).
\bibitem{Fel57} Feller, William \sl An Introduction to Probability Theory 
and its Applications \rm vol I Wiley (1957).
\bibitem{Gra72} Gragg,~W.~B. \sl The Pad\'e Table and its Relation to
Certain Algorithms of Numerical Analysis \rm SIAM Review \bf 14 \rm (1972)
1--62.
\bibitem{Lid97} Lidl,~R. \& Niederreiter,~H. \sl Introduction to Finite
Fields and their Applications \rm Cambridge (1997).
\bibitem{Lot83} Lothaire,~M. \sl Combinatorics on Words \rm Addison-Wesley (1983).
\bibitem{Lun01} Lunnon,~W.~F. \sl The Number-Wall Algorithm: an LFSR Cookbook
\rm Article 01.1.1 Journal of Integer Sequences \bf 4 \rm (2001).
\bibitem{Min67} Minsky,~M. \sl Computation: Finite and Infinite Machines
\rm Prentice-Hall (1967).
\bibitem{Ste92} Stephens,~N.~M. \sl The Zero-square Algorithm for
Computing Linear Complexity Profiles \rm in Mitchell, Chris (ed.)
\sl Cryptography and Coding II \rm Clarendon press Oxford (1992) 259--272.

\end{thebibliography}

\end{document}